\documentclass[fleqn,twoside]{article}
\usepackage[headings]{stylefile}

\readRCS
$Id: stylefile.tex,v 1.2 2004/02/24 11:22:11 spepping Exp $

\usepackage{graphicx}
\usepackage[figuresleft]{rotating}

\newcommand{\AmS}{{\protect\the\textfont2
  A\kern-.1667em\lower.5ex\hbox{M}\kern-.125emS}}

\title{Forward detectors around the CMS interaction point at LHC and their physics potential}

\author{Monika Grothe\address[WI]{Department of Physics, University of Wisconsin \\ 
        1150 University Avenue, Madison, WI 53706, USA}%
        \thanks{monika.grothe@cern.ch}
         $\!\!$, on behalf of the CMS collaboration }
       
\runtitle{Forward detectors around the CMS interaction point at LHC and their physics potential}
\runauthor{M. Grothe}

\begin{document}

\begin{abstract}
Forward physics with CMS at the LHC covers a wide range of physics subjects, including 
very low-$x_{Bj}$ QCD, underlying event and multiple interactions characteristics, 
$\gamma$-mediated processes, shower development at the energy scale of primary cosmic ray 
interactions with the atmosphere, diffraction in the presence of a hard scale and even 
MSSM Higgs discovery in central exclusive production. We describe the forward 
detector instrumentation around the CMS interaction point and present selected 
feasibility studies to illustrate their physics potential.
\vspace{1pc}
\end{abstract}

\maketitle

\section{FORWARD PHYSICS AND FORWARD INSTRUMENTATION AT THE CMS INTERACTION POINT}

Forward physics at the LHC covers a wide range of diverse physics subjects which have in 
common that particles produced at small polar angles $\theta$ and hence large 
values of rapidity provide a defining characteristics. This article concentrates
on their physics interest in $pp$ collisions.

At the Large-Hadron-Collider (LHC), where proton-proton collisions occur at 
center-of-mass energies of 14 TeV, the maximal possible rapidity is 
$y_{max} = \ln{\frac{\sqrt{s}}{m_{\pi}}}\sim 11.5$. The two multi-purpose detectors ATLAS and 
CMS at the LHC are designed primarily for efficient detection of processes with large
polar angles and hence
high transverse momentum $p_T$. The coverage in pseudorapidity 
$\eta = - \ln{[\tan{( \theta / 2 )} ] }$ of
their main components extends down to about $|\theta| = 1^\circ$ from the beam axis
or $|\eta| = 5$.

For the CMS detector, several subdetectors with coverage beyond $|\eta| =5$ are 
currently under construction (CASTOR and ZDC sampling calorimeters) or in the proposal 
stage (FP420 proton taggers and fast timing detectors). 

Futhermore, a salient feature of the forward instrumentation around the 
interaction point
(IP) of CMS is the presence of TOTEM~\cite{TOTEM}. TOTEM is an approved experiment at 
the LHC for measuring the $pp$ elastic cross section as a function of the four-momentum
transfer squared, $t$, and for measuring the total cross section with a precision of
approximately 1\%. The TOTEM experiment uses the same IP as CMS and supplements around 
the CMS IP several tracking devices, located inside of the volume 
of the main CMS detector, plus near-beam proton taggers a distance up to $\pm 220$~m. 
The CMS and TOTEM collaborations have described the considerable physics potential of
joint data taking in a report to the LHCC \cite{opus}. 

The kinematic coverage of the combined CMS and TOTEM apparatus is unprecedented at a
hadron collider. It would be even further enhanced by complementing CMS with the
detectors of the FP420 proposal which would induce forward physics into the portfolio of
possible discovery processes at the LHC~\cite{fp420}.

\begin{figure}
\hspace*{-0.5cm}
\includegraphics[scale=0.32, angle = -90]{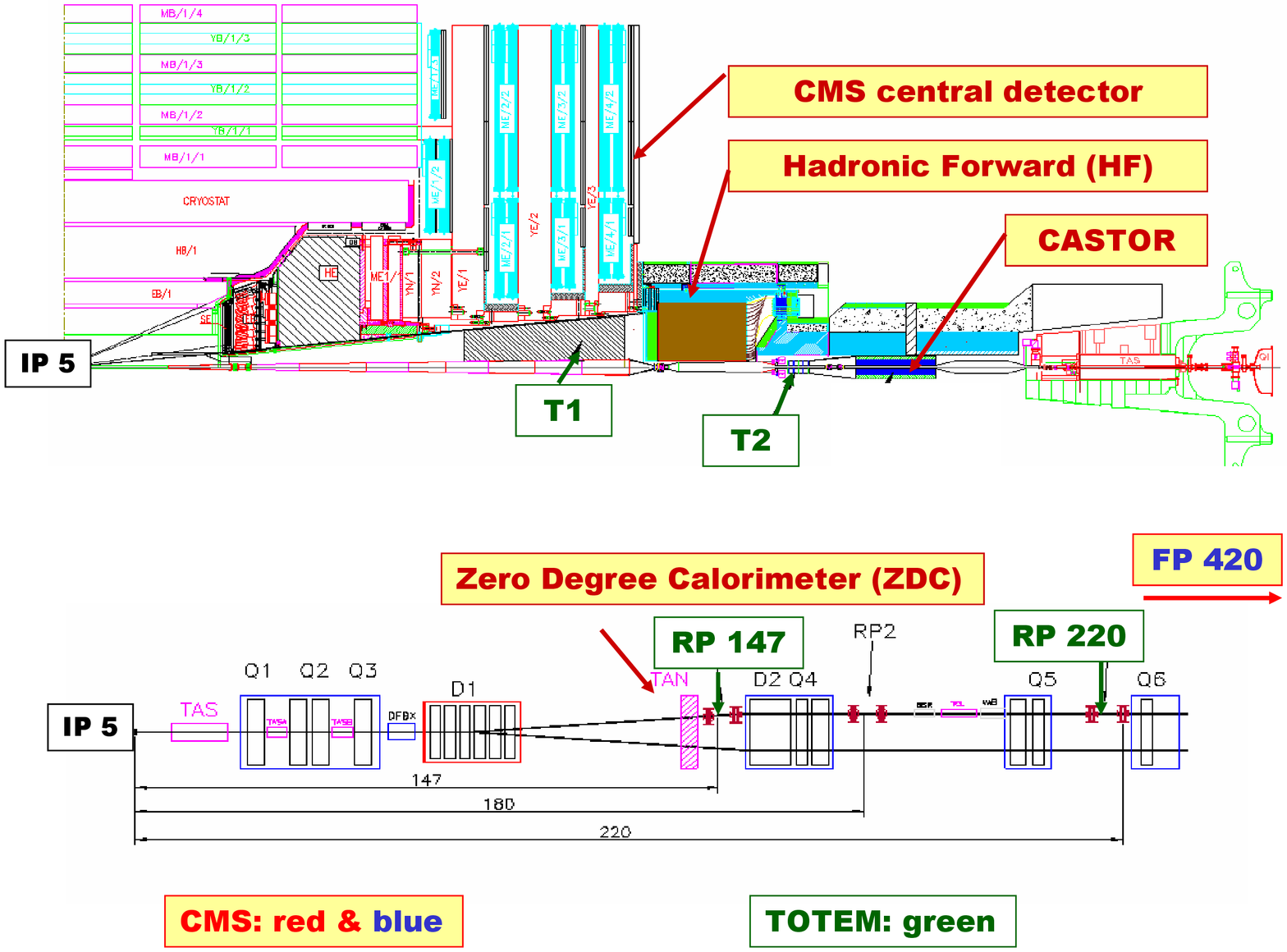}
\caption{Layout of the forward detectors around the CMS interaction point.}
\label{fig:overview}
\end{figure}

An overview of the forward instrumentation up to $\pm 220$~m from the CMS IP is given in 
Fig.~\ref{fig:overview}. There are two suites of calorimeters with tracking detectors in front.
The CMS Hadron Forward (HF) calorimeter with the TOTEM telescope T1 in front 
covers the region $3 < |\eta | < 5$, the CMS CASTOR calorimeter with the TOTEM telescope 
T2 in front covers $5.2 < |\eta| < 6.6$. The CMS ZDC calorimeters will be 
installed at the end of the straight LHC beam-line section, at a distance of 
$\pm 140$~m from the IP. Near-beam proton taggers will be installed by TOTEM at 
$\pm 147$~m and $\pm 220$~m from the IP.
Further near-beam proton taggers in combination with very fast timing detectors to be
installed at $\pm 420$~m from the IP are part of the FP420 proposal.

\section{PHYSICS WITH FORWARD DETECTORS}

In the following, we describe the physics interest of the CMS CASTOR and ZDC 
calorimeters~\cite{PTDR1} 
and the TOTEM T1 and T2 telescopes~\cite{TOTEM}. Of particular interest are 
QCD measurements at values of Bjorken-$x$ as low as $x \sim 10^{-6}$ and the resulting 
sensitivity to non-DGLAP dynamics, 
as well as forward particle and energy flow measurements. These can play an important 
role in tuning the Monte Carlos description of underlying event and multiple interactions
at the LHC and in constraining Monte Carlo generators used for cosmic ray studies.

\subsection{CMS CASTOR \& ZDC calorimeters}

The two calorimeters are of interest for $pp$, $pA$ and $AA$ running at the LHC, 
where $A$ denotes a heavy ion. They are Cherenkov-light devices with electromagnetic 
and hadronic sections and will be present in 
the first LHC $pp$ runs at luminosities where event pile-up should be low.

The CASTOR calorimeters are octagonal cylinders located at $\sim 14$~m from the IP.
They are sampling calorimeters with tungsten plates as absorbers and fused silica quartz 
plates as active medium. The plates are inclined by $45^\circ$ with respect to the 
beam axis. Particles passing through the quartz emit Cherenkov
photons which are transmitted to photomultiplier tubes through aircore lightguides.
The electromagnetic section is 22 radiation lengths $X_0$ deep
with 2 tungsten-quartz sandwiches, the hadronic section consists of 12 tungsten-quartz
sandwiches. The total depth is 10.3 interaction lengths $\lambda_l$. The calorimeters
are read out segmented azimuthally in 16 segments and logitudinally in 14 segments. 
They do not have any segmentation in $\eta$. The CASTOR coverage of 
$5.2 < |\eta| < 6.6$ closes hermetically the total CMS calorimetric pseudorapidity range over
13 units. 

Currently, funding is available only for a CASTOR calorimeter on one side of the IP.
Construction is advanced, with concluding beamtests foreseen for this summer and 
installation in time for the 2009 LHC data taking. 

The CMS Zero Degree Calorimeters, ZDC, are located inside the TAN absorbers 
at the ends of the straight section of 
the LHC beamline, between the LHC beampipes, at $\pm 140$~m distance on each side of the 
IP. They are very radiation-hard sampling calorimeters 
with tungsten plates as absobers and as active medium quartz fibers read out via
aircore light guides and photomultiplier tubes.
The electromagnetic part, $19 X_0$ deep, is segmented into 5 units horizontally, the 
hadronic part into 4 units in depth. The total depth is 6.5 $\lambda_l$. The ZDC 
calorimeters have 100\% acceptance for neutral particles with $|\eta|>8.4$ and can measure
50~GeV photons with an energy resolution of about 10\%. 

The ZDC calorimeters are already installed and will be operational already in 2008.

\subsection{TOTEM T1 \& T2 telescopes}

The TOTEM T1 telescope consists of two arms symmetrically installed around the CMS IP 
in the endcaps of the
CMS magnet, right in front of the CMS HF calorimeters and with $\eta$ coverage similar to
HF.
Each arm consists of 5 planes of Cathod Strip Chambers (CSC) which measure
3 projections per plane, resulting in a spatial resolution of 0.36~mm in the radial and
0.62~mm in the azimuthal coordinate in test beam measurements.

The two arms of the TOTEM T2 telescope are mounted right in front of the CASTOR 
calorimeters, with similar $\eta$ coverage. Each arm consists of 10 planes of 20
semi-circular modules of Gas Electron Multipliers (GEMs). The detector read-out is
organized in strips and pads, a resolution of $115~\mu $m for the radial coordinate and
of $16~\mu$rad in azimuthal angle were reached in prototype test beam measurements.

\subsection{Proton-proton collisions at low $x_{Bj}$}

In order to arrive at parton-parton interactions at very low $x_{Bj}$ values, several 
steps in the QCD cascade initiated by the partons from the 
proton may occur before the final hard interaction takes place. Low-$x_{Bj}$ QCD hence offers
ideal conditions for studying the QCD parton evolution dynamics. Measurements at the
HERA $ep$ collider have explored low-$x_{Bj}$ dynamics down to values of a few $10^{-5}$.
At the LHC the minimum accessible $x$ decreases by a factor $\sim 10$ for each
2 units of rapidity. A process with a hard scale of $Q ~ 10$~GeV and within the 
acceptance of T2/CASTOR ($\eta = 6$) can occur at $x$ values as low as 
$10^{-6}$.

\begin{figure}[htb]
\includegraphics[scale =0.4]{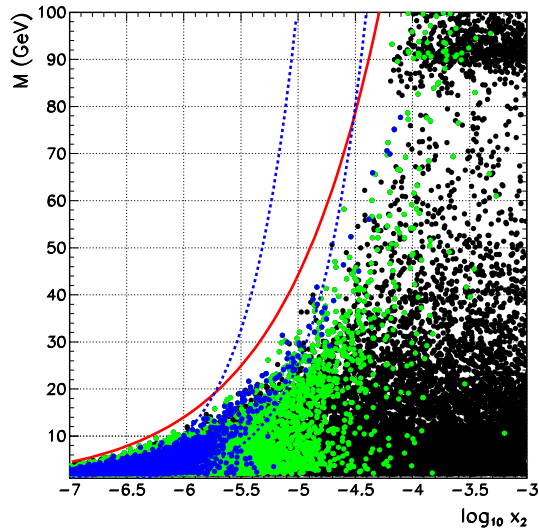}
\caption{Acceptance of the T2/CASTOR detectors for Drell-Yan electrons, see text.}
\label{fig:DYcoverage}
\end{figure}

Forward particles at the LHC can be produced in collisions between two partons with
$x_1 >> x_2$, in which case the hard interaction system is boosted forward.
An example is Drell-Yan production of $e^+ e^-$ pairs, 
$ q q \rightarrow \gamma^\star \rightarrow e^+ e^-$, a process that probes primarily 
the quark content of the proton. Figure~\ref{fig:DYcoverage} shows the distribution of the invariant
mass $M$ of the $e^+ e^-$ system versus the $x_{Bj}$ of one of the quarks, where
$x_2$ is chosen such that $x_1 >> x_2$. The solid curve shows the kinematic limit
$M^{max} = \sqrt{x_2 s}$. The dotted lines indicate the acceptance window for both
electrons to be detectable in T2/CASTOR.
The black points correspond to any of the Drell-Yan events generated 
with Pythia, the green/light grey (blue/dark grey ) ones refer to those events in which at least one (both)
electron lies within the T2/CASTOR detector acceptance. For invariant masses of the $e^+ e^-$ 
system of $M> 10$~GeV, $x_{Bj}$ values down to $10^{-6}$ are accessible.

\begin{figure}[htb]
\includegraphics[scale =0.3]{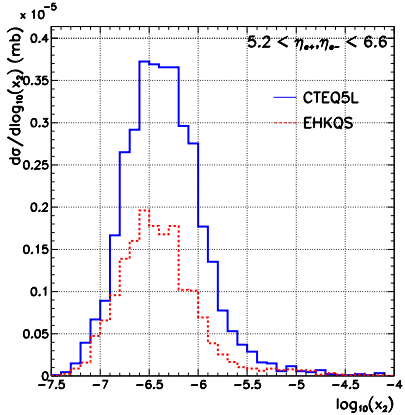}
\caption{Comparison of the cross section prediction of a model without (CTEQ5L) and 
with (EHKQS) saturation for Drell-Yan events in which both electrons are detected in T2/CASTOR.}
\label{fig:saturation}
\end{figure}

The rapid rise of the gluon density in the proton with decreasing values of $x_{Bj}$
observed by HERA in deep inelastic scattering cannot continue indefinitely without violating
unitarity at some point. Hence, parton recombination within the proton must set in at low
enough values of $x_{Bj}$ and leads to non-linear terms in the QCD gluon evolution. 
Figure~\ref{fig:saturation} compares for 
Drell Yan processes with both electrons within the T2/CASTOR detector acceptance the cross
section predicted by a PDF model without (CTEQ5L~\cite{CTEQ}) and with (EHKQS~\cite{EHKQS}) 
saturation effects. A difference of a factor 2 is visible in the predictions. Further details 
can be found in~\cite{opus}.

Complementary information on the QCD evolution at low $x_{Bj}$ can be gained from
forward jets. The DGLAP evolution~\cite{DGLAP} 
assumes that parton emission in the cascade is strongly 
ordered in transverse momentum while in the BFKL evolution~\cite{BFKL}, 
no ordering in $k_t$ is assumed,
but strong ordering in $x$. At small $x_{Bj}$, the difference between the two approaches is
expected to be most pronounced for hard partons created at the beginning of the cascade, 
at pseudorapidities close to the proton, i.e. in the forward direction. Monte Carlo generator
studies indicate that the resulting excess of forward jets with high $p_T$, observed
at HERA, might be measurable with T2/CASTOR. Another observable sensitive to
BFKL-like QCD evolution dynamics are dijets with large rapidity separation, which 
enhances the available phase space for BFKL-like parton radiation between the jets.
Likewise dijets separated by a large rapidity gap are of interest since they indicate
a process in which no color flow occurs in the hard scatter but where, contrary to the 
traditional picture of soft Pomeron exchange, also a high transverse momentum transfer 
occurs across the gap. 

\subsection{Multiplicity \& energy flow}

The forward detectors can be valuable tools for Monte Carlo tuning.

The hard scatter in hadron-hadron collisions takes place in a dynamic environment,
refered to as the ``underlying event'' (UE), where
additional soft or hard interactions between the partons and 
initial and final state radiation occur. The effect of the UE can not be disentangled on an
event-by-event basis and needs to be included by means of tuning Monte Carlo multiplicities 
and energy flow predictions to data. The predictive power of these tunes obtained 
from Tevatron data is very limited, and ways need to be found to constrain the UE at LHC 
energies with LHC data. As shown in~\cite{Borras}, the forward detectors are sensitive
to features of the UE which central detector information alone cannot constrain.

\begin{figure}[!b]
\includegraphics[scale =0.55]{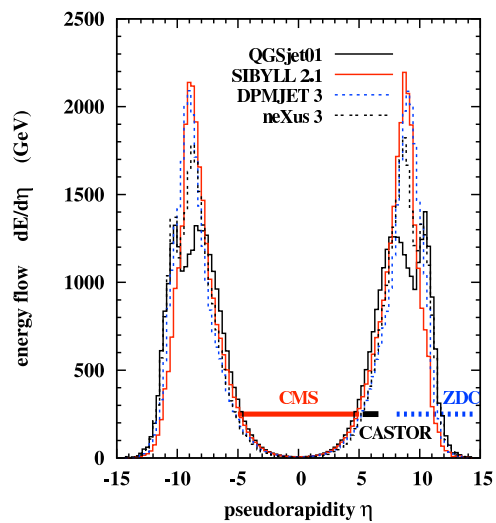}
\caption{Energy flow as predicted by Monte Carlo generators used for the description of 
cosmic ray induced air showers~\cite{opus}.}
\label{fig:cosmics}
\end{figure}

Another area with high uncertainties is modelling the interaction of primary cosmic rays in 
the PeV energy range with the atmosphere. Their rate of occurance per year is too low for
reliable quantitative analysis. The center-of-mass energy in $pp$ collisions at the LHC 
corresponds to 100 PeV energy in a fixed target collision. Figure~\ref{fig:cosmics} shows the 
energy flow
as function of pseudorapidity as predicted by different Monte Carlos in use in the cosmic ray
community. Clear differences in the predictions are visible in the acceptance region of
T2/CASTOR and ZDC.

\section{PHYSICS WITH A VETO ON FORWARD DETECTORS}

Events of the type $pp \rightarrow pXp$ or $pp \rightarrow Xp$, where no color exchange
takes place between the proton(s) and the system $X$, can be caused by $\gamma$ exchange,
or by diffractive interactions. In both cases, the absence of color flow between the
proton(s) and the system $X$ results in a large gap in the rapidity distribution of the
hadronic final state. Such a gap can be detected by requiring the absence of a signal in 
the forward detectors. In the following, we discuss three exemplary processes which are 
characterized by a large rapidity gap in their hadronic final state.

\subsection{Diffraction with a hard scale}

Diffraction, traditionally thought of as soft process and described in Regge theory, can also
occur with a hard scale ($W$, dijets, heavy flavors) as
has been experimentally observed at UA8, HERA and Tevatron. In the presence of a hard scale,
diffractive processes can be described in perturbative QCD (pQCD) and their cross sections
can be factorized into that one of the hard scatter and a diffractive particle 
distribution function (dPDF). In diffractive hadron-hadron scattering, rescattering between
spectator particles breaks the factorization. The so-called rapidity gap survival 
probability quantifies this effect~\cite{survival}. A measure for it can be obtained by the ratio of
diffractive to inclusive processes with the same hard scale. At the Tevatron, the ratio 
is found to be ${\cal O}(1 \%)$~\cite{tevatron}.
Theoretical expectations for the LHC vary from a fraction of a percent to as much as 
30\%~\cite{predLHC}. 

Single diffractive $W$ production, $pp \rightarrow pX$, where $X$ includes a $W$, 
is an example for diffraction with a hard scale at the LHC and is in 
particular sensitive to the quark component of the proton dPDF in an as-of-yet unmeasured 
region. In the absence of event pile-up, a selection is possible based on the requirement
that there be no activity above noise level in the CMS forward calorimeters HF and CASTOR.

\begin{figure}
\includegraphics[scale=0.4]{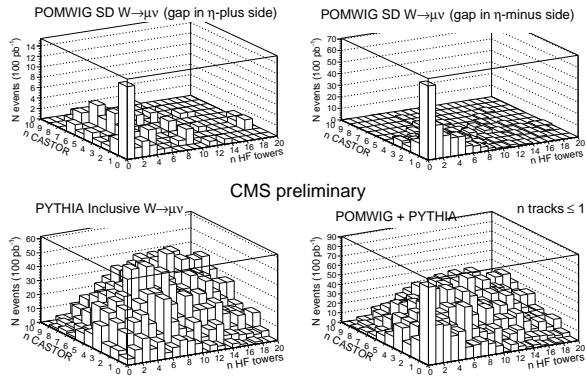}
\caption{Number of towers with activity above noise level in HF versus in CASTOR for
single diffractive $W$ production and for an integrated luminosity of 100~${\rm pb}^{-1}$~\cite{SDW}}
\label{fig:SDW}
\end{figure}

Figure~\ref{fig:SDW} shows the number of towers with activity above noise level in HF versus 
in CASTOR. The decay channel is $W \rightarrow \mu \nu$ and a rapidity gap survical factor 
of 5\% is assumed in the diffractive Monte Carlo sample (Pomwig). The number of events is
normalized to an integrated luminosity of 100~$\rm pb^{-1}$ of single interactions (i.e. no
event pile-up). In the combined Pomwig + Pythia Monte Carlo sample, a clear excess in the 
bin [n(Castor), n(HF)] = [0,0] is visible, of ${\cal O}(100)$ events. The ratio of diffraction 
to non-diffraction in the [0,0] bin of approximately 20 demonstrate the feasibility of 
observing single diffractive $W$ production at the LHC.

The study assumes that CASTOR will be available only on one side. A second CASTOR in the 
opposite hemisphere and the use of T1, T2 will improve the observable excess 
further. 

\subsection{Exclusive dilepton production}

Exclusive dimuon and dielectron production with no significant additional
activity in the CMS detector occurs with high cross section in
gamma-mediated processes at the LHC, either as the pure QED process 
$\gamma \gamma \rightarrow ll$ 
or in $\Upsilon$ photoproduction~\footnote{Photoproduction of $J/psi$ mesons is also 
possible, but difficult to observe because of the trigger thresholds for leptons in CMS.}  
A feasibility study to detect them with CMS was presented in this
workshop~\cite{Hollar}. 

The event selection is based on requiring that outside of the two leptons, no other 
significant activity is visible within the central CMS detector, neither in the calorimeter
nor in the tracking system. In 100 $\rm pb^{-1}$ of single interaction data, ${\cal O} (700)$
events in the dimuon channels and ${\cal O} (70)$ in the dielectron channel can be selected.
Events in which one of the  protons in the process does not stay intact but dissociates 
are the dominant source of background and are comparable in statistics to the signal. 
This background can be significantly reduced by means of a veto condition on activity in CASTOR
and ZDC, in a configuration with a ZDC on each side and a CASTOR on only one side of the IP
by 2/3.

The theoretically very precisely known cross section of the almost pure QED process 
$pp \rightarrow pllp$ via $\gamma$ exchange is an ideal calibration channel. With
$100 \rm pb^{-1}$ of data, an absolute luminosity calibration with 4\% precision is feasible.
Futhermore, exclusive dimuon production is an ideal alignment channel with high statistics 
for the proposed proton taggers at 420~m from the IP. Upsilon photoproduction can constrain 
QCD models of diffraction, as discussed in the next section. 
The $\gamma \gamma \rightarrow e^+ e^-$ process has recently been observed at the Tevatron~\cite{exclTevatron}.

\subsection{Upsilon photoproduction} 

Assuming the STARLIGHT~\cite{starlight} 
Monte Carlo cross section prediction, the 1S, 2S and 3S resonances
will be clearly visible in $100 \rm pb^{-1}$ of single interaction data. With their average
$\gamma p$ center-of-mass energy of $<W_{\gamma p}> \simeq 2400 \rm GeV^2$ they will extend
the accessible range of the HERA measurement of the $W_{\gamma p}$ dependence of 
$\sigma (\gamma p \rightarrow \Upsilon(1 S) p)$ by one order of magnitude.

\begin{figure}[htb]
\includegraphics[scale=0.4]{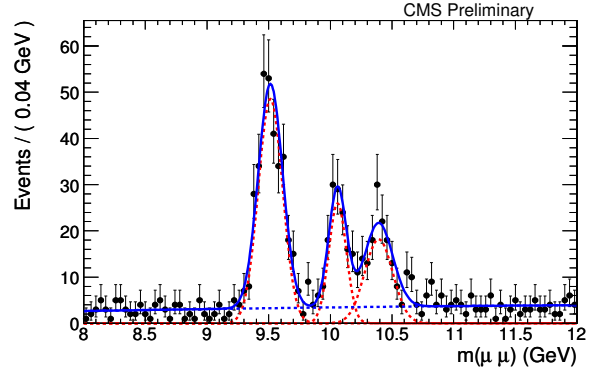}
\caption{Invariant mass of exclusive dimuon production in the Upsilon mass region~\cite{ExclDileptons}}
\label{fig:Upsilon}
\end{figure}

By means of the $p_T^2$ value of the $\Upsilon$ as estimator of the transfered four-momentum 
squared, $t$, at the proton vertex, it might be possible to measure the $t$ dependence of the 
cross section. This dependence is sensitive to the two-dimensional gluon distribution of the 
proton and would give access to the generalized parton distribution function (GPD) of the 
proton.

\section{PHYSICS WITH NEAR-BEAM PROTON TAGGERS}

For slightly off-momentum protons, the LHC beamline with its magnets is essentially a
spectrometer. If a scattered proton is bent sufficiently, but little enough to remain within 
the beam-pipe, they can be detected by means of detectors inserted into the beam-pipe and
approaching the beam envelope as closely as possible. At ligh luminosity at the LHC,
large rapidity gaps typical for diffractive events or events with $\gamma$ exchange tend to be 
filled in by particles from overlaid pile-up events. Hence tagging the outgoing scattered 
proton(s) becomes the only mean of detection at high luminosities.

\subsection{TOTEM and FP420 proton taggers}

The TOTEM proton taggers, located at $\pm 147$~m and $\pm 220$~m from the IP, each consist
of Silicon strip detectors housed in movable Roman Pots~\cite{TOTEM}. 
The detector
design is such that the beam can be approached up to a minimal distance of $10 \sigma +$~0.5~mm.
With nominal LHC beam optics, scattered protons from the IP are within the acceptance of the 
taggers at 220~m when for their fractional momentum loss $\xi$ holds: $0.02 < \xi < 0.2$.

\begin{figure}
\includegraphics[scale=0.35, angle =-90]{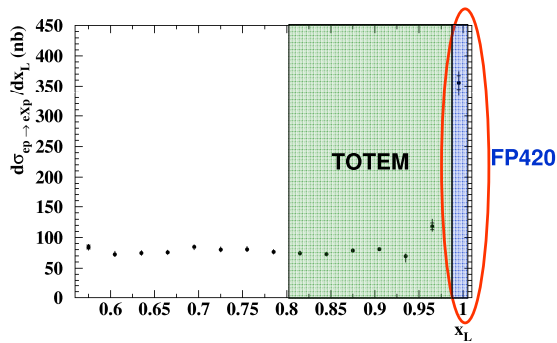}
\caption{Acceptance in $x_L = 1 - \xi$, where $\xi$ is the fractional momentum loss of the 
scattered proton, of the TOTEM and FP420 proton taggers. The data points shown are from 
ZEUS~\cite{zeus}.}
\label{fig:xiCoverage}
\end{figure}

In order to achieve acceptance at smaller values of $\xi$ with nominal LHC beam optics, 
detectors have to be located further away from the IP. Proton taggers at $\pm 420$~m from the
IP have an acceptance of $0.002 < \xi < 0.02$, complementing taggers at 220~m, as shown
in Figure~\ref{fig:xiCoverage}. 
The proposal~\cite{fp420} of the FP420 R\&D collaboration foresees employing 3-D Silicon, an
extremely radiation hard novel Silicon technology, for the proton taggers. Additional 
fast timing Cherenkov detectors will be capable of determining, within a resolution of a 
few millimeters, whether the tagged proton came from the same vertex as the hard scatter visible
in the central CMS detector. In order to comply with the space constraints of the location 
within the cryogenic region of the LHC, these detectors will be attached to a movable beam-pipe
with the help of which the detectors can approach the beam to within 3~mm.

The FP420 proposal is currently under scrutiny in CMS and ATLAS. If approved, installation could
proceed in 2010, after the LHC start-up.

\subsection{Physics potential}

Forward proton tagging capabilities enhance the physics potential of CMS. They would
render possible a precise measurement of the mass and quantum numbers of the Higgs boson
should it be discovered by traditional searches. They also augment the CMS discovery reach
for Higgs production in the minimal supersymmetric extension (MSSM) of the Standard Model (SM)
and for physics beyond the SM in $\gamma p$ and $\gamma \gamma$ interactions.

A case in point is the central exclusive production (CEP) process~\cite{CEP}, 
$pp \rightarrow p + \phi + p$, where the plus sign denotes the absence of hadronic 
activity between the outgoing protons, which survive the interaction intact, and the 
state $\phi$. The final state consists solely of the
scattered protons, which may be detected in the forward proton taggers, and the decay 
products of $\phi$ which can be detected in the central CMS detector. 
Selection rules force the produced state $\phi$ to have $J^{CP} = n^{++}$ with $n =0, 2, ..$. 
This process offers hence an experimentally very clean 
laboratory for the discovery of any particle with these quantum numbers that couples 
strongly to gluons. Additional advantages are the possibility to determine the mass of the state
$\phi$ with excellent resolution from the scattered protons alone, independent of its
decay products, and the possibility, unique at the LHC, to determine the quantum numbers of 
$\phi$ directly from the azimuthal asymmetry between the scattered protons.

\begin{figure}[!b]
\includegraphics[angle=-90]{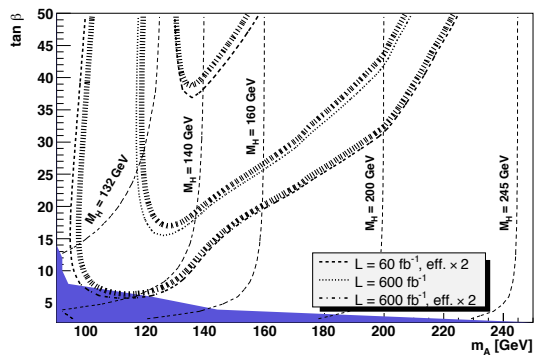}
\caption{Five $\sigma$ discovery contours for central exclusive production of the 
heavier CP-even Higgs boson $H$~\cite{Tasevsky}. See text for details.}
\label{fig:higgs}
\end{figure}

In the case of a SM Higgs boson with mass close to the current exclusion limit, which decays
preferably into $b \bar{b}$, CEP
improves the achievable signal-to-background ratio dramatically, to 
$\cal{O}$(1)~\cite{fp420,lightHiggs}. 
In certain
regions of the MSSM, generally known as ``LHC wedge region'', the heavy MSSM Higgs bosons would 
escape detection at the LHC. 
There, the preferred search channels at the LHC are not available 
because the 
heavy Higgs bosons decouple from gauge bosons while their couplings to $b \bar{b}$ and 
$\tau \bar{\tau}$ are enhanced at high $\tan{\beta}$. Figure~\ref{fig:higgs} depicts
the 5~$\sigma$ discovery contour for the $H \rightarrow b \bar{b}$ channel in CEP in 
the $M_A - \tan{\beta}$ plane of the MSSM within the $M_h^{max}$ benchmark scenario
with $\mu = +200$~GeV and for different integrated luminosities. 
The values of the mass of the heavier CP-even Higgs boson, $M_H$, are indicated by 
contour lines. The dark region corresponds to the parameter region excluded by LEP. 

Forward proton tagging will also give access to a rich QCD program on hard diffraction
at high luminosities, where event pile-up is significant and makes undetectable the gaps 
in the hadronic final state otherwise typical of diffraction. Detailed studies with high
statistical precision will be possible on skewed, unintegrated gluon 
densities; Generalized Parton Distributions which contain information on the correlations 
between partons in the proton; and the rapidity gap survival probability, a quantity closely 
linked to soft rescattering effects and the features of the underlying event at the LHC.

Forward proton tagging also provides the possibility for precision studies of $\gamma p$
and $\gamma \gamma$ interactions at center-of-mass energies never reached before. Anomalous top
production, anomalous gauge boson couplings, exclusive dilepton production and quarkonia 
production are possible topics, as was discussed in detail at this workshop.

\section{SUMMARY}

Forward physics in $pp$ collisions at the LHC covers a wide range of diverse physics subjects (low-$x_{Bj}$ QCD,
hard diffraction, $\gamma \gamma$ and $\gamma p$ interactions)
 which have in
common that particles produced at large
values of rapidity provide a defining characteristics. 
For the CMS detector, several subdetectors with forward $\eta$ coverage 
are currently under construction (CASTOR, ZDC) or in the proposal 
stage (FP420). The TOTEM experiment 
supplements around the CMS IP several tracking devices and near-beam proton taggers at 
distances up to $\pm 220$~m. 
The kinematic coverage of the combined CMS and TOTEM apparatus is unprecedented at a
hadron collider. It would be even further enhanced by complementing CMS with the
detectors of the FP420 proposal which would add forward physics to the portfolio of
possible discovery processes at the LHC.

\end{document}